\documentstyle[12pt]{article}

\newcommand{\mysection}{\setcounter{equation}{0}\section}

\begin{document}
\hfill{FSU-HEP-951030} 
\vskip 0.1cm
\hfill{ITP-SB-95-15} 
\vskip 0.1cm
\hfill{LBL-37266} 
\vskip 0.3cm
\centerline{\large\bf {Reanalysis of the EMC charm production data with 
extrinsic}}
\centerline{\large\bf {and intrinsic charm at NLO}}
\vskip 0.3cm
\centerline {\sc B. W. Harris}
\centerline{\it Physics Department,} 
\centerline{\it Florida State University,}
\centerline{\it Tallahassee, Florida 32306-3016 USA}
\vskip 0.3cm
\centerline{\sc J. Smith}
\centerline{\it Institute for Theoretical Physics,}
\centerline{\it State University of New York at Stony Brook,}
\centerline{\it Stony Brook, New York 11794-3840 USA}
\vskip 0.3cm 
\centerline{and}
\vskip 0.3cm 
\centerline{\sc R. Vogt\footnote{This work was supported in part by
the Director, Office of Energy Research, Division of Nuclear Physics
of the Office of High Energy and Nuclear Physics of the U. S.
Department of Energy under Contract Number DE-AC03-76SF0098.}}
\centerline{\it Nuclear Science Division,}
\centerline{\it Lawrence Berkeley Laboratory,}
\centerline{\it Berkeley, California 94720 USA} 
\centerline{\it and}
\centerline{\it Physics Department,}
\centerline{\it University of California at Davis,}
\centerline{\it Davis, California 95616 USA}
\vskip 0.3cm
\centerline{November 1995}
\vskip 0.5cm
\centerline{\bf Abstract}
\vskip 0.3cm

A calculation of the next-to-leading order exclusive
extrinsic charm quark differential 
distributions in deeply inelastic electroproduction has 
recently been completed.  
Using these results we compare the NLO extrinsic contributions to
the charm structure function $F_2(x,Q^2,m_c^2)$ with the corresponding 
NLO intrinsic contributions.  
The results of this analysis are compared with the EMC DIS charm 
quark data and evidence for an intrinsic charm component in the proton 
is found.

\vfill
\newpage
\mysection{Introduction}
\newcommand{\be}{\begin{eqnarray}}
\newcommand{\ee}{\end{eqnarray}}

The European Muon Collaboration
(EMC) \cite{emc} has played an important role in the study of charmed quark
production in deep inelastic scattering.  Their results established
that photon-gluon fusion in QCD, the analogue of
the Bethe-Heitler reaction in QED,  explains most of the charmed quark
contribution to the deep inelastic structure function $F_2$.  
The scattering cross section for the reaction 
$e^-(l_1) + P(p) \rightarrow e^-(l_2) + c(p_1) + X$ 
is conventionally written as a differential in $x$ and $y$ as
\be
\frac{d^2\sigma}{dxdy} = \frac{2\pi\alpha^2}{Q^4} S 
\left[ \left\{ 1 + (1-y)^2 \right\} 
F_2(x,Q^2,m_c^2) - y^2 F_L(x,Q^2,m_c^2) \right]
\ee
with $q=l_1-l_2, Q^2=-q^2, x = Q^2/2p\cdot q$ and $y = p\cdot q/p\cdot l_1$.  
The dependence on the charm quark mass, $m_c$, is shown explicitly.  
The square of the center-of-mass energy of the 
electron-proton system is $S$.  The above formula holds after 
integration over the azimuthal angle between the plane containing
the incoming and outgoing electron and the plane containing the incoming
proton and outgoing charmed quark.  
In a typical experimental extraction of $F_2$ from $\sigma$, the $F_L$ 
contribution is either neglected or approximated by leading order QCD.

The lowest order (LO) photon-gluon fusion model in QCD
is based on the twist-two term in the operator product 
expansion which incorporates the factorization theorem \cite{CSS2} for
hard scattering.
The EMC $F_2(x,Q^2,m_c^2)$ data was not in complete agreement with 
the predictions of this model at all $x$ and $Q^2$. 
The disagreement at large $x$ and $Q^2$ substantiated the claim that a 
second (higher twist) 
component of charm production was necessary, called
intrinsic charm (IC) \cite{intc} 
to distinguish it from the twist-two mechanism, 
referred to as extrinsic charm (EC).

In the analysis of the EMC data \cite{emc}, 
a relatively simple model was used for
both the EC and IC components.  At the time, only the LO contributions 
to $F_2(x,Q^2,m_c^2)$ from both models were available. 
Afterwards, Hoffman and Moore \cite{hm} calculated the 
next-to-leading order (NLO) corrections to
the IC component and discussed their effects on
the EMC analysis.  Based on LO photon-gluon fusion, they found evidence for an
0.3 \% IC component in the proton.  
Due to the lack of NLO calculations of the EC component
for the analysis of the EMC data, the
previous results are inconclusive.

The NLO corrections to the 
extrinsic component are now also available 
\cite{LRSvN12,RSvN2,LRSvN22,hs1,hs2}. 
These NLO calculations yield large cross sections in specific regions of phase
space since new mechanisms occur, such as the $t$-channel 
exchange of massless gluons.  Hence cross sections for ``massive'' quark
production in NLO should really not be compared with the LO prediction
via the standard $K$ factor, 
although this is often done in the literature.  We will comment on this later.

The single particle inclusive results \cite{LRSvN22} 
for the charmed quark distributions
were applied to the EMC data in \cite{grs} without addressing the IC component.
However, even the single particle inclusive NLO results 
are not enough for a complete 
reanalysis of the EMC data since the invariant mass of the 
$c \overline{c}$ pair, $M_{c \overline{c}}$, was included in the scale of the
running coupling constant and cuts were made on 
the energies of the decay muons. Thus an exclusive NLO calculation,
retaining all possible information on both the charmed and the 
anticharmed quarks, is required.  The NLO corrections to exclusive
production were derived in \cite{hs1,hs2}, making it possible to include
$M_{c\bar c}$ in the running coupling constant. Since the complete NLO results
are now available for both EC and IC production,
a more detailed QCD analysis of the EMC results is finally possible,
allowing us to make a more reliable determination of the IC content 
of the proton.

To make this paper self-contained, we begin with a
review of intrinsic charm and its NLO contribution to the 
structure function $F_2(x,Q^2,m_c^2)$ in Section 2.
We follow this with a brief review of extrinsic charm and its NLO 
contribution to $F_2(x,Q^2,m_c^2)$ in Section 3. 
In Section 4 we discuss the results of our analysis  
and conclude that, even with the addition of the new NLO EC 
contributions and the use of the most recent parton densities,
an IC component is still needed to fit the EMC data.

\mysection{The Intrinsic Charm Component}

The QCD wavefunction of a hadron can be represented as a
superposition of quark and gluon Fock states. For example, at fixed
light-cone time, $\tau= t + z/c$, the proton wavefunction can be
expanded as a sum over the complete basis of free quark and gluon
states: $\vert \Psi_p \rangle = \sum_m \vert m \rangle \,
\psi_{m/p}(x_i, k_{T,i}, \lambda_i)$ where the color-singlet
states, $\vert m \rangle$, represent the fluctuations in the proton
wavefunction with the Fock components $\vert
u  u  d \rangle$, $\vert u u d g \rangle$, $\vert
u u d c \overline c \rangle$, {\it etc}. The boost-invariant
light-cone wavefunctions, $\psi_{m/p}(x_i, k_{T,i}, \lambda_i)$,
needed to compute probability distributions, are functions of the
relative momentum coordinates $x_i = k_i^+/P^+$ and $k_{T,i}$.
Momentum conservation demands $\sum_{i=1}^n x_i = 1$ and
$\sum_{i=1}^n \vec{k}_{T,i}=0$, where $n$ is the number of partons
in a Fock state $\vert m \rangle$.  When an interaction occurs, 
the coherence of the Fock components is broken and the
fluctuations can hadronize, forming new hadronic systems \cite{BHMT}. 
For example, intrinsic $c \overline c$
fluctuations \cite{intc} can be liberated provided the system is probed during
the characteristic time, $\Delta t = 2p_{\rm lab}/M^2_{c \overline
c}$, that such fluctuations exist.

There is substantial circumstantial evidence for
the existence of intrinsic $c
\overline c$ states.
 Leading charm production in $\pi N$ \cite{769} and hyperon-$N$ \cite{WA89}
collisions requires a charm source beyond leading twist.  The leading
charm production can be explained by the coalescence of intrinsic charmed
quarks in the projectile wavefunction with spectator valence quarks
\cite{VB,VBH2}.  Final state coalescence mechanisms \cite{PYT}
cannot consistently describe data with meson and baryon projectiles.  
The NA3 experiment has shown that the
single $J/\psi$ cross section at large $x_F$ is greater than
expected from $gg$ and $q \overline q$ production \cite{Bad,VBH1}. 
This experiment
has also measured a significant number of correlated $J/\psi$ pairs 
\cite{Badpp}
which carry a larger fraction of the projectile momentum than expected from
leading-twist QCD.  Intrinsic heavy quark states such as $|\overline u d c
\overline c c \overline c \rangle$ can explain the production of
fast $J/\psi$ pairs \cite{VB2}.
Additionally, intrinsic charm may account for the anomalous
longitudinal polarization of the $J/\psi$
at large $x_F$ \cite{Van} seen in $\pi N \to J/\psi X $ interactions.

Microscopically, the intrinsic heavy quark Fock component in the
proton wavefunction, $|u u d c \overline c \rangle$, is
generated by virtual interactions such as $g g \rightarrow Q
\overline Q$ where the gluons couple to two or more 
valence quarks. The probability for $c \overline c$ fluctuations to
exist in a light hadron thus scales as $\alpha_s^2(m_c^2)/m_c^2$
relative to leading-twist production \cite{VB}. Therefore, this
contribution is higher twist,
suppressed by ${\cal O}(1/m_c^2)$ compared to extrinsic production.

The dominant Fock state configurations are not far off
shell and thus have minimal invariant mass, $M^2 = \sum_i m_{T,
i}^2/ x_i$ where $m_{T, i} = \sqrt{k^2_{T,i}+m^2_i}$ is the
transverse mass of parton $i$ in the state.
Intrinsic $c \overline c$ Fock components with minimum invariant
mass correspond to configurations with equal rapidity constituents.
Thus, unlike extrinsic heavy quarks generated from a single parton, intrinsic
heavy quarks carry a larger fraction of the parent momentum
than the light quarks in the state \cite{intc}. It was shown that
large $x_F$ virtual $c\overline c$ or lepton pairs can be liberated
by a relatively soft interaction \cite{BHMT}. 
For soft interactions at momentum scale
$\mu_s$, the intrinsic heavy quark cross section is suppressed by a
resolving factor proportional to $\mu_s^2/4 m^2_c$ \cite{VB} for 
hadroproduction and $(\mu_s^2+Q^2)/(4m_c^2+Q^2)$ for electroproduction 
\cite{BHMT,chev}.

The general form of the Fock state wavefunction appropriate to any frame at
fixed light-cone time is \be
\Psi(\vec k_{\perp i},x_i) = \frac{\Gamma(\vec k_{\perp i},x_i) }{m_h^2 -
M^2 } \, \, , \ee where $\Gamma$ is a
vertex function, expected to be a decreasing function of $m_h^2 - M^2$, a 
measure of the ``off-shellness" of the Fock state fluctuation. 
The vertex function is assumed to be
relatively slowly varying; the particle distributions are then
controlled by the light-cone energy denominator and  phase space.
This form for the higher Fock components is applicable to an
arbitrary number of light and heavy partons. 
The Fock states containing charmed quarks can be materialized
by a soft collision in the target which brings the state
on shell. In the limit of zero binding energy, $\Psi$ is singular and
the fractional momenta peak at $x_i = m_i/m_h$.  Note that the denominator is
minimized when the heaviest constituents carry the largest fraction of the
longitudinal momentum.  The parton distributions
reflect the underlying shape of the Fock state wavefunction.
Assuming it is sufficient to
use a mean value of $k_T^2$ to calculate the $x$ distributions,
the probability distribution for a general $n$--particle intrinsic
$c \overline c$ Fock state as a function of $x$ is
\be
\frac{dP_{\rm ic}}{dx_i \cdots dx_n} =  N_n [\alpha_s^2(M_{c
\overline c})]^2
\ \frac{\delta(1-\sum_{i=1}^n x_i)}{(m_h^2 - \sum_{i=1}^n
(\widehat{m}_i^2/x_i)
)^2} \, \, ,
\ee
where $\widehat{m}_i = \sqrt{m_i^2 + \langle
\vec k_{T, i}^2 \rangle}$ is the average transverse mass and 
$N_n$ normalizes the Fock state probability.
In the heavy quark limit, $\widehat{m}_c$, $\widehat{m}_{\overline c} \gg m_h$,
$\widehat{m}_q$ and the probability reduces to \be 
\frac{dP_{\rm ic}}{dx_i \cdots dx_n} =  N_n [\alpha_s^2(M_{c
\overline c})]^2 \frac{x_c^2 x_{\overline c}^2}{(x_c + x_{\overline c})^2}
\ \delta(1-\sum_{i=1}^n x_i) \, \, . \ee  

Integration over the light quarks
and anticharmed quark momenta for $n=5$, the minimal intrinsic charm Fock 
state of the proton, $|uudc \overline c \rangle$, gives the intrinsic charmed
quark density distribution as a function of the charmed quark momentum 
fraction,
\be
\label{cdef}
c(x) \propto \frac{dP_{\rm ic}(x)}{dx} = 
\frac{1}{2} N_5 x^2 [\frac{1}{3}(1-x)(1 +
10x + x^2) + 2x(1+x)\ln x] \, \, . 
\ee  
If there is a 1\% probability for intrinsic charm in the nucleon, as 
previously suggested \cite{intc}, then $N_5 = 36$.  The charmed quark 
structure function at leading order $F_2^{(0)}(x)$ is given by 
\be
\label{massless}
F_2^{(0)}(x) = 8 x c(x) /9 \,,
\ee
in the limit where the charmed quark mass is negligible.  In a complete 
analysis of the EMC charm data, the massless result is clearly inapplicable.  

Hoffmann and Moore \cite{hm} incorporated mass effects into the above 
analysis.  They first introduced a mass scaling 
variable $\xi = 2ax [1 + (1 + 4 \rho x^2 )^{1/2}]^{-1} $ where 
$\rho = m_p^2/Q^2$, $a = [(1 + 4 \lambda)^{1/2} +1]/2$ and 
$\lambda = m_c^2/Q^2$.  The proton mass is denoted by $m_p$.  The
$c \overline c$ mass threshold imposes the 
constraint $\xi \leq \gamma < 1$ where 
$\gamma = 2a\hat x [1 + (1 + 4\rho \hat x^2)^{1/2}]^{-1}$.
Then (\ref{massless}) is replaced by
\be
F_2^{(0)}(x,Q^2,m_c^2) = 8 \xi c(\xi,\gamma)/9 \, ,
\ee 
with $c(z,\gamma)=c(z)-zc(\gamma)/\gamma$ for $z \leq \gamma$ and zero 
otherwise; $c(z)$ is defined in (\ref{cdef}).
Hoffmann and Moore found that further mass effects could be incorporated 
by generalizing the operator-product expansion analysis 
to include both the charmed quark and target masses. 
The final LO result, {\it c.f.}\ eq.\ (18) in \cite{hm}, is then 
\be 
F_2^{(0)}(x,Q^2,m_c^2)  = \frac{8x^2}{9(1+4\rho x^2)^{3/2}}
\left[ \frac{(1+4 \lambda)}{\xi} c(\xi, \gamma)
+ 3 \hat g (\xi,\gamma) \right] \,,
\ee
where
\be
\hat g(\xi,\gamma) = \frac{2\rho x}{(1 + 4 \rho x^2)}
\int_\xi^\gamma\, dt \frac{c(t,\gamma)}{t}
\left( 1 - \frac{\lambda}{\rho t^2} \right)
\left[1 + 2 \rho x t + \frac{2 \lambda x}{t}\right] \,.
\ee
The NLO IC component of the structure function is given by 
\be
F_2^{(1)} (x, Q^2, m_c^2) = \frac{8}{9} \xi \int_{\xi / \gamma}^1 \frac{dz}{z} 
c(\xi / z, \gamma) \sigma_2^{(1)} (z, \lambda) \,.
\ee
When the lowest order cross section is normalized to 
\be
\sigma^{(0)}_2(z,\lambda) = \delta(1-z),
\ee
the NLO QCD corrections
to the IC contribution are given by eq.\ (51) in \cite{hm}, 
\be
\label{full}
 && \sigma^{(1)}_2(z,\lambda)  =  
\frac{2\alpha_s}{3\pi}
\delta(1-z) \Big\{ 4 \ln\lambda -2 + \sqrt{1 + 4 \lambda} L 
+\frac{(1+2\lambda)}{\sqrt{1+4\lambda}}
[3L^2 
\nonumber \\ && 
+ \, 4L + 4{\rm Li}_2(-d/a)
+ 2L\ln \lambda -4 L \ln(1+4\lambda) + 2{\rm Li}_2(d^2/a^2)]\Big\}
\nonumber \\ &&
+ \, \frac{\alpha_s}{3\pi} \frac{1}{(1+ 4\lambda z^2)^2}
\Big\{\frac{1}{[1- (1-\lambda) z]^2}
\nonumber \\ &&
\times[(1-z)(1-2z-6z^2+8z^4)
+6\lambda z(1-z)(3-15z-2z^2+8z^3)
\nonumber \\ &&
+ \, 4\lambda^2 z^2 (8-77z+65z^2-2z^3)
+16\lambda^3 z^3 (1-21z +12z^2) - 128 \lambda^4 z^5]
\nonumber \\ &&
- \, \frac{2 \hat L}{\sqrt{1 + 4 \lambda z^2}} [(1+z)(1+2z^2)
-2\lambda z (2-11z -11 z^2) -8\lambda^2 z^2 (1-9z)]
\nonumber \\ &&
- \, \frac{8z^4(1+4\lambda)^2}{(1-z)_+}
-\frac{4z^4(1+2\lambda)(1+4\lambda)^2 \hat L}{\sqrt{1+4\lambda z^2} (1-z)_+}
\Big\} \,,  
\ee
where
\be
\hat L = \ln \left[ \frac{ 4 \lambda z[1 - (1-\lambda)z]}
{(1 + 2\lambda z + \sqrt{1 + 4 \lambda z^2} )^2} \right] \,.
\ee
The leading-logarithmic approximation to this
formula, given by eq.\ (54) in the same paper, is
\be
\label{ll}
\tilde\sigma^{(1)}_2(z, \lambda) & = & \frac{\alpha_s}{3\pi} 
\left\{ \frac{(1-2z-6z^2)}{(1-z)_+}
- \frac{2(1+z^2)\ln z}{1-z} \right. \nonumber \\ &&
  \left. -\frac{2(1+z^2)\ln \lambda}{(1-z)_+}
- 2(1+z^2) \Big[ \frac{\ln(1-z)}{1-z}\Big]_+ \right. 
 \nonumber \\ &&
\left.-\delta(1-z)[3\ln \lambda +5 +2\pi^2/3] \right\} \, .
\ee
The implementation of the plus distributions in eqs.\ (\ref{full}) and 
(\ref{ll}) is standard \cite{plus} :  
\be
\int_a^1 dx f(x) \left( \frac{1}{1-x} \right)_{+} = \int_a^1 dx 
\frac{f(x)-f(1)}{1-x} + f(1) \ln(1-a)
\ee
\begin{eqnarray}
\int_a^1 dx f(x) \left( \frac{\ln(1-x)}{1-x} \right)_{+} &=& \int_a^1 dx
\frac{f(x)-f(1)}{1-x} \ln(1-x) \nonumber \\ &+& \frac{1}{2} f(1) \ln^2(1-a).
\ee

In our calculations of the IC contributions to $F_2(x,Q^2,m_c^2)$ we have
assumed a 1\% probability for IC in the proton.  To compare with the original
EMC analysis, 
we take $m_c = 1.5$ GeV and fix the renormalization scale in the one loop
running coupling constant to
$\mu^2 = Q^2 + 20\: {\rm GeV}^2$ and $\Lambda^{(4)}_{QCD} = 0.5$ GeV.  Later,
when the EC calculations are updated with more recent parton densities, we
take the same value of $\Lambda^{(4)}_{QCD}$ as in the parton densities and use
the two loop running coupling constant to be consistent with the EC results.

Figure 1 shows the LO and NLO IC contributions 
to $F_2(x,Q^2,m_c^2)$ for two $Q^2$ values, $Q^2 = 7$ and 70 GeV$^2$.
The upper dotted line shows the massless ($Q^2$ independent) result, (2.5), 
the dot-dashed line shows the $\xi$ scaling formula, (2.6), and the upper
solid line shows the full result, (2.7), all at LO.  At low $Q^2$ there is 
substantial difference between the simple form given in (2.5) and the 
kinematically corrected formula in (2.7).  
The three curves are nearly indistinguishable when $Q^2$ is large,
as seen in Fig.\ 1(b).  In the same figure we also show the 
NLO corrections calculated using eq.\ (2.9).
These are the leading-logarithmic approximation eq.\ (2.13), 
dotted lines, and the complete result, eq.\ (2.11), dashed lines.  
These two formulae yield different curves at small $Q^2$ but
are almost identical at large $Q^2$.  

We note here that we have uncovered an error in the 
implementation of eq.\ (2.9) in recent work \cite{Gol}:
the constant term in eq.\ (2.14), $f(1) \ln(1-a)$, was omitted.  
This term is due to the non-zero lower limit of integration in eq.\ (2.9) 
and has a large numerical effect.  Contrary to \cite{Gol}, 
our results show that the heavy quarks evolve faster than the massless quarks 
using the standard DGLAP equations, as suggested by the original 
results of Hoffmann and Moore \cite{hm}. 

In the remainder of this paper, for the IC results at NLO,  
we use the sum of the kinematically corrected LO result, eq.\ (2.7), 
and the full NLO correction obtained by using eq.\ (2.9) 
with eq.\ (2.11).
The result is given by the lower (at $x=0.4$) solid line in Fig. 1.  
Note that this total is actually negative at large $x$ and $Q^2$,
indicating that even higher order terms are needed in the perturbation 
expansion in this region.

\mysection{The Extrinsic Charm Component}

Order $\alpha_s$ QCD corrections to the charmed quark structure 
functions for inclusive production in deep inelastic scattering 
were first presented in \cite{LRSvN12}.  Heavy quark production 
was assumed to be extrinsic so that $F_2$ and $F_L$ could be 
calculated from the {\em inclusive} virtual photon-induced reaction
$ \gamma^\ast(q) + P(p) \rightarrow  c(p_1) + X$.  The parton level
interaction is 
\begin{eqnarray}
\label{partint}
\gamma^\ast(q) + a_1(k_1) \rightarrow  c(p_1) + \overline c(p_2) + a_2(k_2)
\end{eqnarray}
where $a_1$ and $a_2$ are massless partons and $p_2$ and $k_2$ are integrated
over in the inclusive calculation.
The structure functions can then be written as 
\begin{eqnarray}
\label{fhad}
F_{k}(x,Q^2,m^2_c) &=& \frac{Q^2 \alpha_s(\mu^2)}{4\pi^2 m^2_c} 
\int_{\xi_{\rm min}}^1 \frac{d\xi}{\xi}  \left[ \,e_c^2 f_{g/P}(\xi,\mu^2)
 c^{(0)}_{k,g} \, \right] \nonumber \\ 
&+& \frac{Q^2 \alpha_s^2(\mu^2)}{\pi m^2_c} \int_{\xi_{\rm min}}^1 
\frac{d\xi}{\xi} \left\{ \,e_c^2 f_{g/P}(\xi,\mu^2_c) 
\left( c^{(1)}_{k,g} + \overline c^{(1)}_{k,g} \ln \frac{\mu^2}{m^2_c} \right) 
\right. \nonumber \\ 
&+& \sum_{i=q,\overline q} f_{i/P}(\xi,\mu^2) \left. \left[ e_c^2
\left( c^{(1)}_{k,i} + \overline c^{(1)}_{k,i} \ln \frac{\mu^2}{m^2_c} \right) 
\right. \right. \nonumber \\ 
&+& \left. \left. e^2_i \, d^{(1)}_{k,i} + e_c\, e_i \, o^{(1)}_{k,i} \, 
\Large\right] \Large\right\} \, , \nonumber \\ &&
\end{eqnarray}
where $k = 2,L$.
The lower boundary on the integration is
$\xi_{\rm min} = x(4m^2_c+Q^2)/Q^2$. 
The parton momentum distributions in the proton are denoted by
$f_{i/P}(\xi,\mu^2)$ where $\mu$, 
the mass factorization scale, has been set equal to the 
renormalization scale in the running coupling constant $\alpha_s$.  Finally,  
$c^{(l)}_{k,i}$ and $\overline c^{(1)}_{k,i}\,,
(l=0,1)$,  and $d^{(1)}_{k,i}$ and 
$o^{(1)}_{k,i}$ are scale independent parton coefficient functions. 
In eq.\ (\ref{fhad}) the coefficient functions are distinguished 
by their origin:
$c^{(l)}_{k,i}$ and $ \overline c^{(1)}_{k,i}$ originate from the 
virtual photon-charmed quark coupling and therefore appear for both charged and
neutral parton-induced reactions; $d^{(1)}_{k,i}$ arise from the
virtual photon-light quark coupling; $o^{(1)}_{k,i}$ come 
from the interference between these
processes. All charges are in units of $e$. 
We have included the terms proportional to $e_c e_i$ in (3.2)
even though they do not contribute to the total partonic cross section. 
We have also isolated the mass factorization scale dependent terms,
proportional to $\overline c^{(1)}_{k,i}$.
Finally, note that eq.\ (\ref{fhad}) only holds for $Q^2>0$.  
In the photoproduction limit there are additional terms involving 
the parton densities in the photon \cite{EN,fmnr,SvN2,frix2}.

Recently, two of us have reported on the results of a calculation of the 
NLO corrections for heavy quark {\em exclusive} distributions at 
fixed $x$ and $Q^2$ \cite{hs1,hs2}. 
This allows us to study correlations between the outgoing 
particles in eq.\ (\ref{partint}).
The exclusive results are needed for our reanalysis of the EMC 
data because in the original analysis \cite{emc}, 
the strong coupling constant was evaluated
at the scale $\mu^2 = Q^2 + M_{c \overline c}^2$.
Contrary to what was assumed in previous analysis of the IC component, 
$M_{c \overline c}$ is not a constant but a function of $x$ and $Q^2$ 
\cite{hs2}.

Since the NLO calculations were not available at the time of the EMC
experiment, it is clearly interesting to make a reanalysis of their data 
including both the NLO corrections and more recent parton densities.  
For a complete description at NLO, we have included the quark and 
antiquark contributions to charm production as well as the gluon contribution.
See \cite{hs1} for details. 

Our analysis is done in two steps.  We begin with
the same parameters as used in the EMC analysis \cite{emc} 
to confirm that we can 
reproduce their EC results at LO.  Using these same parameters, we then 
calculate the NLO contribution to see what effect it would have had.  
We have used $m_c = 1.5$ GeV/c$^2$ and the one 
loop running coupling constant at scale
$\mu^2 = Q^2 + M_{c\overline c}^2$ with $\Lambda^{(4)}_{QCD} = 0.5$ 
GeV.  In conjunction with the one loop coupling, the scale independent gluon 
density $x g(x) = 3 (1 - x)^5$ was also used.
We will see that the EC NLO calculation yields values of $F_2(x,Q^2,m_c^2)$
which are above
the data.   This can be corrected by adjusting the normalization of the
theoretical prediction.  

Next we repeat the analysis with
modern parton distributions and the two loop running coupling constant
evaluated at 
$\mu^2=Q^2+M_{c \bar{c}}^2$.  The 
CTEQ3 $\overline{{\rm MS}}$ parton distributions with  
$\Lambda^{(4)}_{QCD}=0.239$ GeV \cite{CTEQ3} are used.  We also compare with
MRS \cite{MRSG} and the most recent GRV \cite{GRV} parton distributions to
investigate the dependence on the parton densities.


\mysection{Results}

We first show the IC and EC results for $F_2(x,Q^2,m_c^2)$ as a function 
of $Q^2$ for fixed $\nu$ and study the differences between the NLO  
and LO results with the running coupling scale and gluon distribution 
used by EMC \cite{emc}.  We retrieved
$F_2(x,Q^2,m_c^2)$ in bins of $x$ and $Q^2$ from the Durham-RAL HEP Database 
\cite{pdg}.  The central $x$ and $Q^2$ values, $\overline{x}$ and 
$\overline{Q^2}$, give the average value of $\nu$,
$\overline{\nu} = \overline{Q^2} / 2m_p \overline{x}$.
The results corresponding to the three curves in Fig.\ 13 of the EMC paper 
are presented in Fig.\ 2 for $\overline{\nu} = 168,\, 95, \,$ and 53 GeV.
Note that the $\nu$ bin widths quoted in \cite{emc} are $160 < \nu < 220$, 
$100 < \nu < 160$, and $60 < \nu < 100$, all in GeV.  At the values of $Q^2$
relevant to the EMC data, the mass of the charmed quark is not negligible.
Thus the charmed quark cannot be treated as massless and absorbed in the parton
densities.

Figure 2 shows the EMC data with our calculations of
the LO EC contribution (dotted lines), the NLO EC 
contribution (solid lines), the LO IC contribution (dot-dashed lines), and 
NLO IC contribution (dashed lines).  
The LO EC contribution agrees with the EMC result, as expected,
since we used their gluon density.
Clearly the EC component can be reduced to improve the agreement 
of the NLO results with the EMC data by changing the
normalization of the gluon density. This also requires some readjustment 
in the normalization of the charged 
parton densities to maintain the momentum sum rule.  In any case, 
the data cannot be reconciled with the EC prediction alone since such a
readjustment will not fit the data at large $\overline \nu$ and $Q^2$.

We next show the EMC data for the structure function $F_2(x,Q^2,m_c^2)$
at $\bar \nu = 53$ GeV in Fig. 3(a) plotted as a function of $x$.
Also in the same figure we add the theory predictions for the LO EC
contribution, the NLO EC contribution, the LO IC contribution
and the NLO IC contribution. Again we have used the EMC gluon density so that
the LO EC result fits most of the data at small $x$ while the IC 
component is important at larger $x$. Figures 3(b) and 3(c) show the
corresponding results for $\bar \nu = 95$ GeV and 168 GeV respectively.
It is again true that if we readjust the normalization of the gluon 
density to move the theory curves down we still cannot
get a good fit to the data using only the EC NLO result.

Rather than elaborate on refitting the old EMC gluon density,
we turn to a comparison of modern gluon densities
with the data. In Fig.\ 4 we again show 
$F_2(x,Q^2,m_c^2)$ for fixed $\bar \nu$ versus $x$
with the corresponding theoretical curves for the CTEQ3 parton densities
\cite{CTEQ3} obtained from global fits to deep-inelastic,
Drell-Yan and direct photon production data. Note that the  
EMC data have not been used in the global analysis to obtain the 
parton densities.   For consistency $m_c = 1.5$ GeV$/c^2$
was used in both the EC and IC contributions.   The two-loop $\alpha_s$ with 
$\Lambda_{\rm QCD}^{(4)} = 0.239 $ GeV 
and matching across mass thresholds was used for the EC results 
which are shown for scales $\mu = \mu_0 /2,
\mu = \mu_0 $ and $\mu = 2 \mu_0$ where
$\mu_0^2 = Q^2 + M_{c\bar c}^2$.
We also show the NLO EC component calculated at $\mu = \mu_0$ 
using the MRS(G) \cite{MRSG}
parton densities with $\Lambda_{\rm QCD}^{(4)} = 0.255 $ GeV 
and the GRV \cite{GRV} parton densities with 
$\Lambda_{\rm QCD}^{(4)} = 0.200 $ GeV. 
All parton densities give essentially the same 
results as there is very little difference between the parton 
densities in this $x$ and $Q^2$ range. 
To demonstrate this we present average values of
$x$ in the $\bar \nu$ bins for all the parton density 
and scale combinations in Table 1.  Finally the IC component 
is shown using the two loop $\alpha_s(\mu)$ 
and the values of $\Lambda_{\rm QCD}$ determined by the parton densities. 
It is clear that the data cannot be fit with only the EC contribution,
even with the most recent parton densities.

We now discuss this conclusion in more detail.  To fit the data
at NLO a considerable enhancement in $F_2(x,Q^2,m_c^2)$ is needed at 
large $x$, significantly beyond that shown in Fig.\ 4.  The $K$ 
factor, $F_2^{NLO}(x,Q^2,m_c^2) / F_2^{LO}(x,Q^2,m_c^2)$, would then 
have to have a strong $x$ dependence.  To illustrate the variations 
between the LO and NLO EC results the $K$ factors, calculated with 
NLO parton densities in the numerator and denominator and with the 
two-loop $\alpha_s$, are shown in Fig.\ 5 as functions of $x$ for the 
EMC $\bar \nu$ values.  We have also computed the $K$ factors with the 
LO parton densties (available for CTEQ3 and GRV) and the lowest 
order coefficient functions with the one-loop $\alpha_s$ in the 
denominator and the NLO parton densities with the two-loop $\alpha_s$ 
in the numerator, as advocated in \cite{grs}.  Although some 
differences in the $K$ factors are observable, they are not 
large enough to affect our conclusions.
We want to remind the reader that the NLO corrections to 
heavy quark production contain dynamical
mechanisms which lead to large enhancements in the cross sections 
\cite{EN,HH,EK} at NLO. 
An analogous situation in QED would be a comparison of the muon pair
production cross section in the reaction $e^+ e^- \rightarrow \mu^+ \mu^-$ 
in LO, where there is an $s$-channel pole, with the NLO
muon pair production by the two photon mechanism, 
$e^+ e^- \rightarrow e^+ e^- \mu^+ \mu^-$, which has a $t$-channel pole.
At large energies, the two photon mechanism clearly yields a larger 
cross section but this does not violate the convergence of
the perturbative expansion for QED.
Therefore it is unwise to place too much emphasis on a 
comparison with the LO results.

The increase in the $K$ factor appears at 
large $Q^2$ and
large $x$, close to threshold, as may be expected \cite{LRSvN22}.  
However, the most 
important issue is whether or not the $K$ factor is a constant in 
the region of the EMC data.  A constant $K$ factor would lead to a uniform
enhancement of the EC component without improving the fit at large $x$.
Indeed, the $K$ factors change by 10-30\% for the higher $\bar{\nu}$ 
bins at large
$x$, not enough to significantly affect the results.  In the lowest 
$\bar{\nu}$ 
bin, the $K$ factor increases by nearly a factor of two, but the EC results
are steeply falling and are negligible compared to the IC contribution for
$x > 0.3$.  Therefore, in the $x$ region covered by EMC, we see 
that the $K$ factor does not change enough to affect our conclusions 
regarding the 
need for an IC component.

In Fig.\ 6 we show the 
corresponding $K$ factors for the IC component.
The opposite behavior to the EC component is observed--the $K$ factor falls 
with $x$, flattening at large $x$.  This is also clearly a threshold 
effect since the largest $K$ factors appear at the highest $\bar{\nu}$ 
and lowest $x$.  In this
region, close to threshold, the IC component is negligible compared to the
EC contribution, thus the large $K$ factor will not affect our conclusions.  
Where the IC and EC components are comparable, the IC
$K$ factor is small.  Note also that the IC $K$ factor becomes less than unity
for $x > 0.2$ depending on $\bar \nu$, as can be anticipated from
the Fig.\ 1.

Now let us try to get a good fit to the data
with the sum of an EC and an IC component to the structure function.
We have performed a least 
squares fit to the data using the Levenberg-Marquardt 
algorithm \cite{bev}.  The normalization 
of both the IC and EC components are taken as free parameters,  
\be 
F_2^c(x,Q^2,m_c^2) = \alpha \cdot F_2^{c, {\rm EC}}(x,Q^2,m_c^2) 
                   + \beta \cdot F_2^{c, {\rm IC}}(x,Q^2,m_c^2) \, \, , 
\ee
with the scale $\mu = \mu_0$.
The shift in the normalization 
of the EC component may be considered as an estimate of the
size of the NNLO contribution, which is equivalent to a shift in
the scale $\mu$.
Since we have already assumed a 1\% normalization of the IC component, the
fitted $\beta$ is the fraction of this normalization.
The results are presented in Table 2.
The errors quoted in the table correspond to 
a $95\%$ confidence level on the central fit parameters.  
The final results for the combined model of eq.\
(4.1) are shown in Fig.\ 7  
for the CTEQ3, MRS(G) and GRV94 sets of parton densities.
The table shows that given the quality of the data, no 
statement can be made about the intrinsic charm content for 
$\bar{\nu}=53$ and $95 \: {\rm GeV}$. However, 
for $\bar{\nu}=168 \: {\rm GeV}$ 
an intrinsic charm contribution of $(0.86\pm0.60)\%$ is indicated.
For completeness we have examined the influence of a 
resolving factor, $R(Q^2) = (\mu_s^2 + Q^2)/(4m_c^2 + Q^2)$ where 
$\mu_s^2 = 0.2 \: {\rm GeV}^2$ for the IC contribution. This does
not alter our conclusions.  

We close with some remarks on other recent results. 
We have already mentioned that another group
\cite{grs} also checked that recent parton densities fit the
EMC data, using slightly older GRV parton densities \cite{grv}.  They
concentrated on the $x$ and $Q^2$ regions where there is clearly
no IC component in an attempt to use $F_2(x,Q^2,m_c^2)$ for
a direct determination of the gluon density in the proton \cite{Sch2}.
Therefore there is no overlap with our work.
A recent analysis of muon production from IC decays 
at HERA (and at fixed target facilities) suggests that even an 0.1\% IC
contribution could be measurable at HERA \cite{ijn}. We therefore urge 
experimentalists to make a decisive measurement.

{\bf Acknowledgements}

The work of B.W. Harris was supported in part under the
contracts NSF 93-09888 and DOE-FG05-87ER40319.
The work of J. Smith was supported in part under the
contract NSF 93-09888.  We thank Yu.A. Golubkov 
for discussions and for providing us with his computer code
and J.F. Owens for discussions.


%

\newpage
\centerline{\bf \large{Figure Captions}}

\begin{description}
\item[Fig.\ 1]
(a) The IC contributions to the structure function $F_2(x,Q^2,m_c^2 )$
at $Q^2 = 7$ GeV: the massless result (2.5) (upper dotted line), 
the $\xi$-scaling result (2.6) (dot-dashed line) 
and the kinematically corrected formula (2.7) (top solid line). 
Also shown are the NLO corrections given by (2.9), 
with the leading-log result (2.13) (dashed line), 
and the full cross result (2.11) (lower dotted line). 
The sum of (2.7) and (2.9) using (2.11),
represents the total IC contribution to $F_2(x,Q^2,m_c^2)$ 
(lower solid line). (b) Same as part (a) for $Q^2=70$ GeV.
\item[Fig.\ 2]
(a) The EC and IC contributions to the proton structure function 
$F_2(x,Q^2,m_c^2)$ plotted as functions of $Q^2$
for fixed $\bar{\nu} = 53$ GeV.  The curves show the predictions from the 
EC photon-gluon fusion model (LO: dotted and NLO: solid lines)
and from a $1\%$ IC component ( LO: dot-dashed and NLO: dashed lines). 
The data are from the EMC experiment \cite{emc} via 
Durham-RAL HEP data base \cite{pdg}. 
(b) The same as (a) for $\bar{\nu} = 95$ GeV.
(c) The same as (a) for $\bar{\nu} = 168$ GeV.
\item[Fig.\ 3]
(a) The EMC data for the structure function $F_2(x,Q^2,m_c^2)$ at 
$\bar{\nu} = 53$ GeV 
plotted as a function of $x$ together with the EC and IC LO and NLO 
contributions.  The notation for the lines is the same as in Fig.\ 2.
(b) Same as (a) for $\bar{\nu} = 95$ GeV.
(c) Same as (a) for $\bar{\nu} = 168$ GeV.
\item[Fig.\ 4]
(a) The EMC data for $F_2(x,Q^2,m_c^2)$ at $\bar \nu= 53$ GeV
plotted as a function of $x$ together with the EC and IC results.
The solid lines are the NLO EC results with 
$\mu= \mu_0/2$ (upper), $\mu= \mu_0$ (middle), $\mu= 2\mu_0$ (lower). 
The dotted line is the NLO result for the MRS(G) parton densities
with $\mu = \mu_0$, while the dashed line is the result for the 
GRV94 parton densities with $\mu = \mu_0$. (b) Same as (a) for 
$\bar{\nu}=95$ GeV. (c) Same as (a) for $\bar{\nu}=168$ GeV.
\item[Fig.\ 5]
(a) The $K$ factors for EC production as a function of $x$ for
$\bar \nu = 53$ GeV. The solid lines are for the CTEQ3 parton densities with
$\mu = 2\mu_0$ (upper), $\mu = \mu_0$ (middle) and $\mu = \mu/2$ (lower).
The dotted lines is for the MRS(G) parton densities 
with $\mu =\mu_0$ and the dashed line is for the GRV94 parton
densities with $\mu = \mu_0$.
(b) The same as (a) for $\bar \nu = 95$ GeV.
(c) The same as (a) for $\bar \nu = 168$ GeV.
\item[Fig.\ 6]
The $K$ factors for IC production as a function of $x$ for
$\bar \nu = 168$ GeV (solid line), $\bar \nu = 95$ GeV 
(dashed line) and $\bar \nu = 53$ GeV (dotted line).
\item[Fig.\ 7]
(a) The EMC data for the structure function $F_2(x,Q^2,m_c^2)$ at
$\bar{\nu} = 53$ GeV plotted as a function of $x$ together with 
the fitted results from (4.1). The solid line is for the 
CTEQ3 parton densities, the dotted line is for the MRS(G) parton densities
and the dashed line is for the GRV94 parton densities.
The parameters $\alpha$ and $\beta$ are given in Table 1.
(b) Same as (a) for $\bar{\nu} = 95$ GeV.
(c) Same as (a) for $\bar{\nu} = 168$ GeV.
\end{description}

\centerline{\bf \large{Table Caption}}
\begin{description}
\item[Table 1]
Average $x$ values of the EC component for $\bar \nu$ bins and the 
various parton densities used in this analysis.
\end{description}
%
%
\begin{description}
\item[Table 2]
Results of the least squares fit of EC and IC contributions to the EMC data  
according to (4.1).  Uncertainties in the fit parameters are shown at the 
$95\%$ confidence level.

\end{description}

\begin{table}
\begin{center}
\begin{tabular}{||c||c|c|c|c||}
\hline
\multicolumn{2}{|c|}{} & \multicolumn{3}{c||}{$ \langle x \rangle $} \\
\hline
PDF & $\mu$  & $\bar{\nu}=53$ GeV & $\bar{\nu}=95$ GeV & $\bar{\nu}=168 $ GeV
  \\
\hline \hline
CTEQ3 &  $\mu_0 /2$ & 0.12 & 0.12 & 0.11 \\
CTEQ3 &  $  \mu_0 $ & 0.12 & 0.11 & 0.10 \\
CTEQ3 &  $2 \mu_0 $ & 0.12 & 0.11 & 0.10 \\
MRS(G) &   $\mu_0   $ & 0.12 & 0.12 & 0.11 \\
GRV(94) &  $\mu_0   $ & 0.13 & 0.12 & 0.11 \\
\hline
\end{tabular}
\end{center}
\caption{}
\end{table}

\begin{table}
\begin{small}
\begin{tabular}{||c||c|c||c|c||c|c||}
\hline
\multicolumn{1}{||c||}{} & \multicolumn{2}{c||}{$\bar{\nu}=53$ GeV} & 
\multicolumn{2}{c||}{$\bar{\nu}=95$ GeV} & \multicolumn{2}{c||}{$\bar{\nu}=168$
GeV} \\
\hline
PDF & $\alpha$ & $\beta$ & $\alpha$ & $\beta$ & $\alpha$ & $\beta$ \\
\hline \hline
CTEQ3 & 0.95 $\pm$ 0.64 & 0.36 $\pm$ 0.58 & 1.20 $\pm$ 0.13 & 0.39 $\pm$ 0.31 
& 1.27 $\pm$ 0.06 & 0.92 $\pm$ 0.53 \\
MRS(G) & 1.02 $\pm$ 0.69 & 0.34 $\pm$ 0.58 & 1.38 $\pm$ 0.15 & 0.32 $\pm$ 0.32
& 1.47 $\pm$ 0.07 & 0.79 $\pm$ 0.53 \\
GRV(94) & 1.15 $\pm$ 0.77 & 0.33 $\pm$ 0.58 & 1.45 $\pm$ 0.16 & 0.34 $\pm$ 0.31
& 1.48 $\pm$ 0.08 & 0.88 $\pm$ 0.53 \\
\hline
\end{tabular}
\end{small}
\caption{}
\end{table}

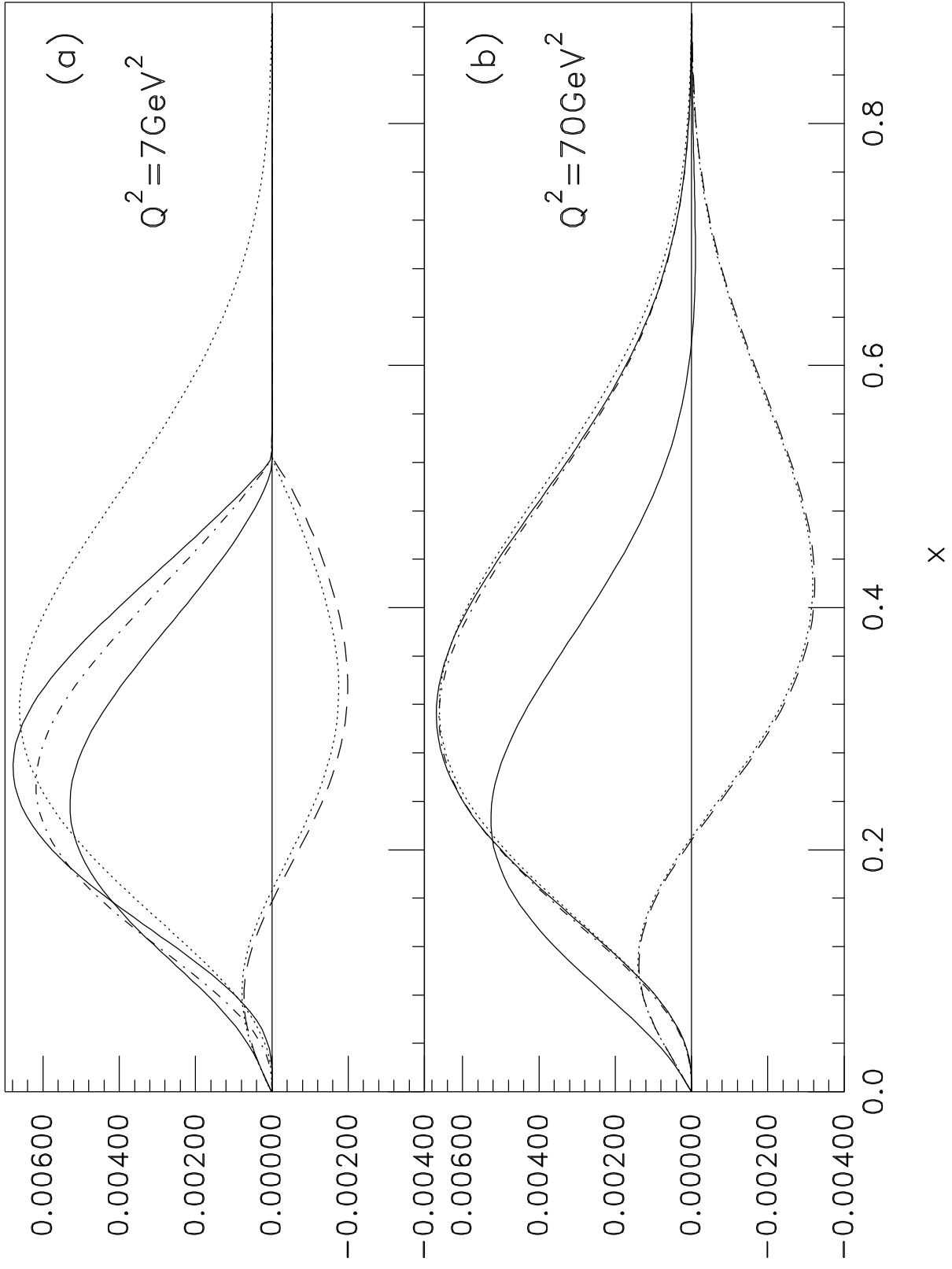
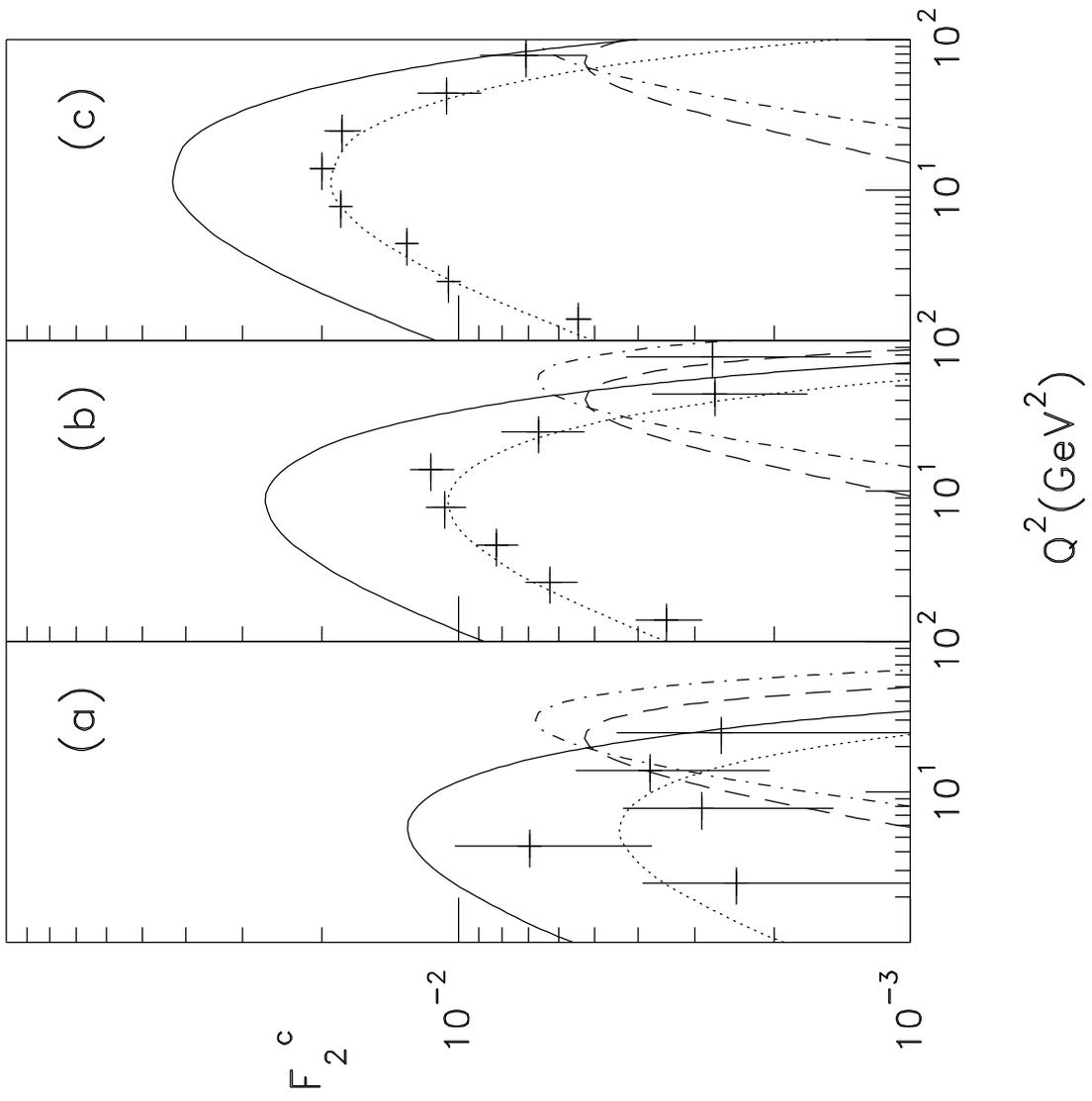
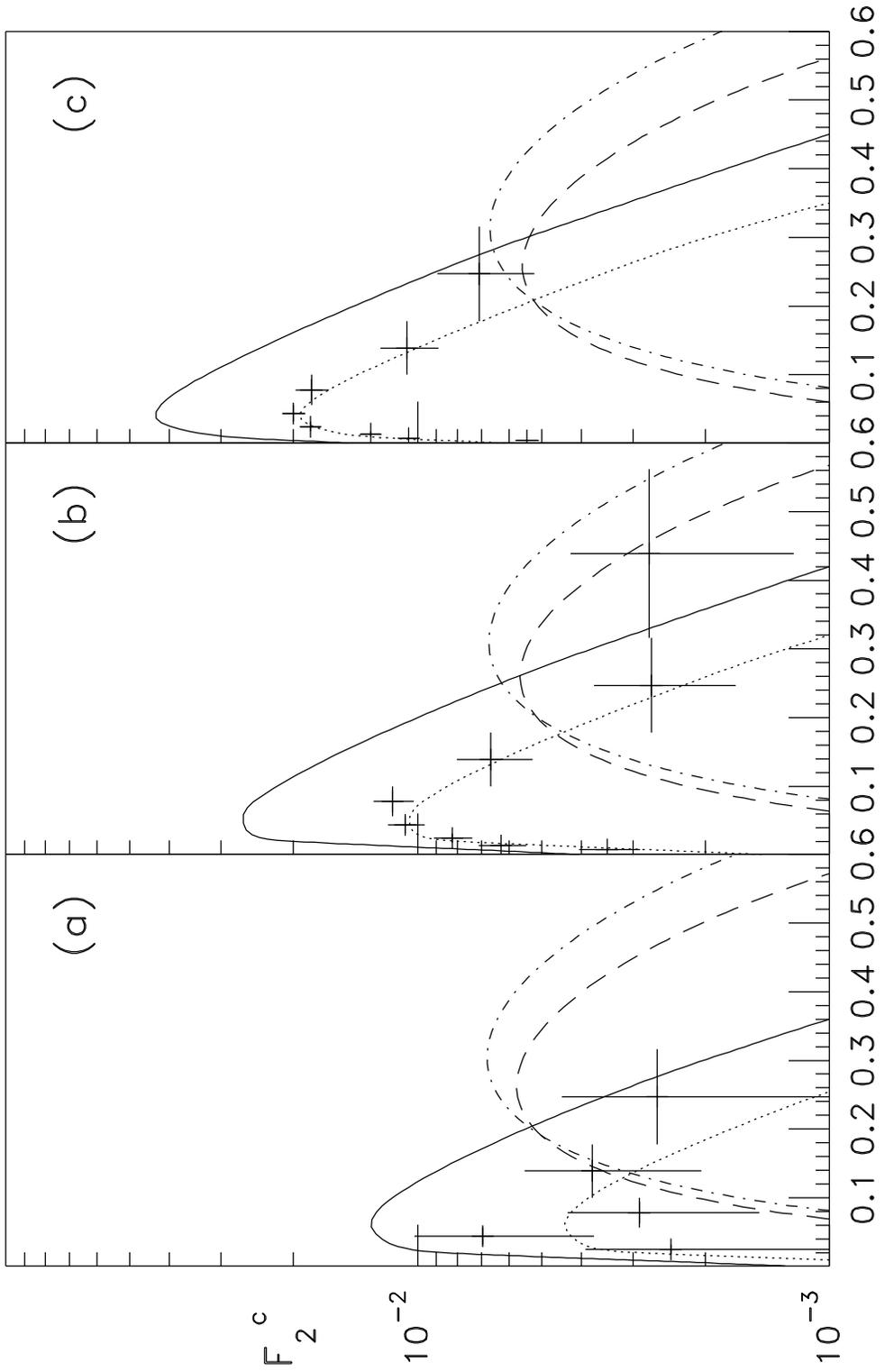
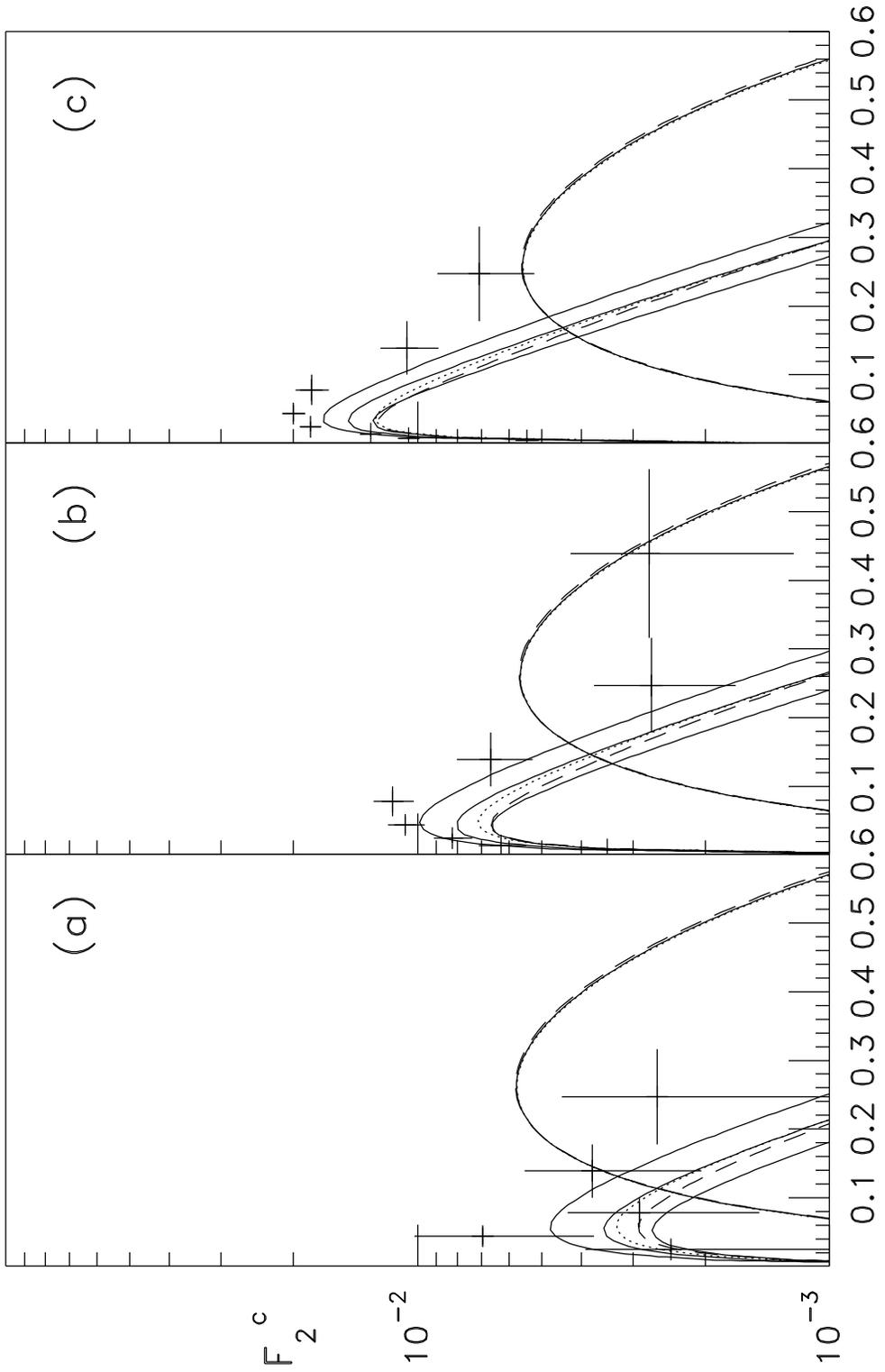
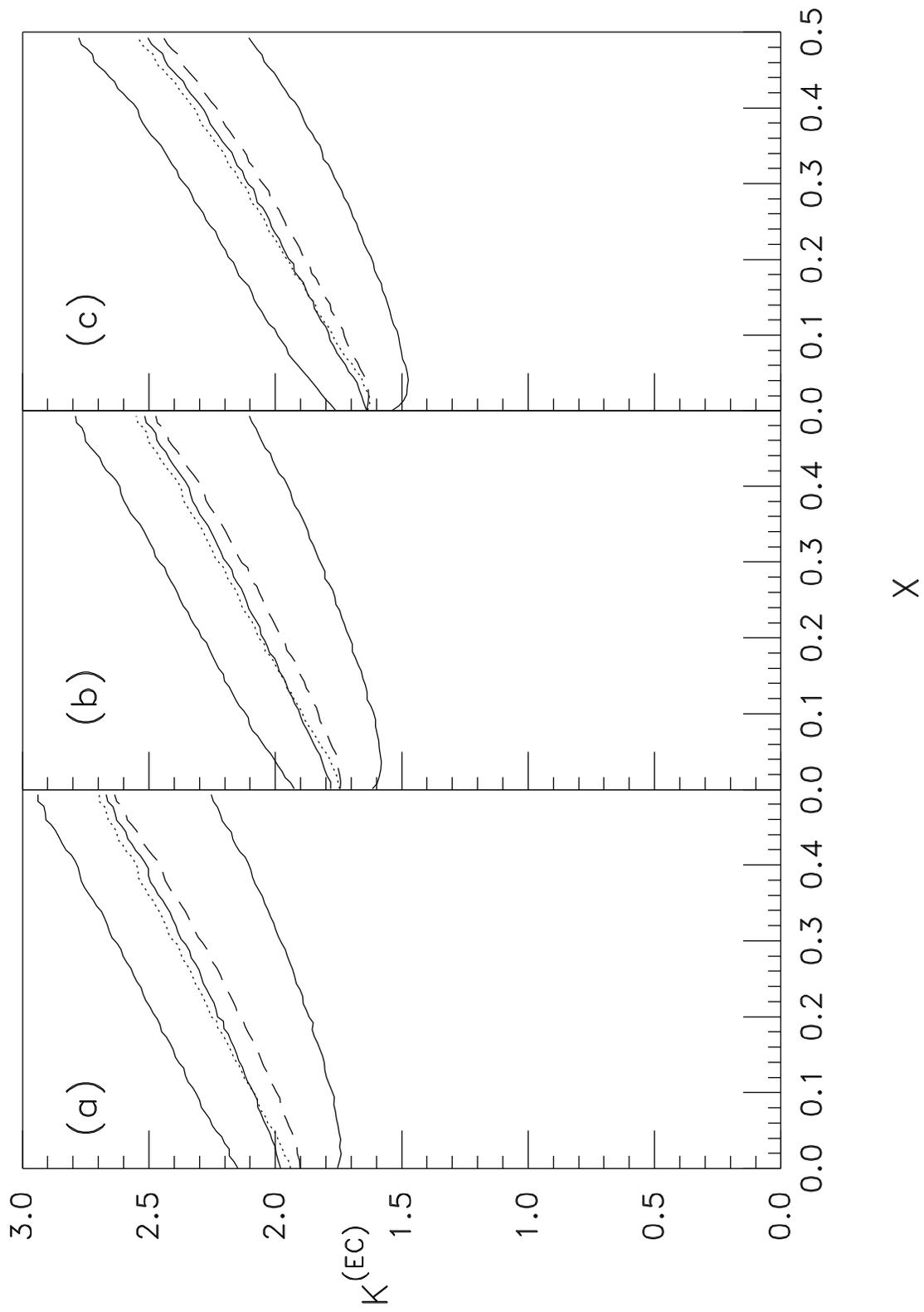
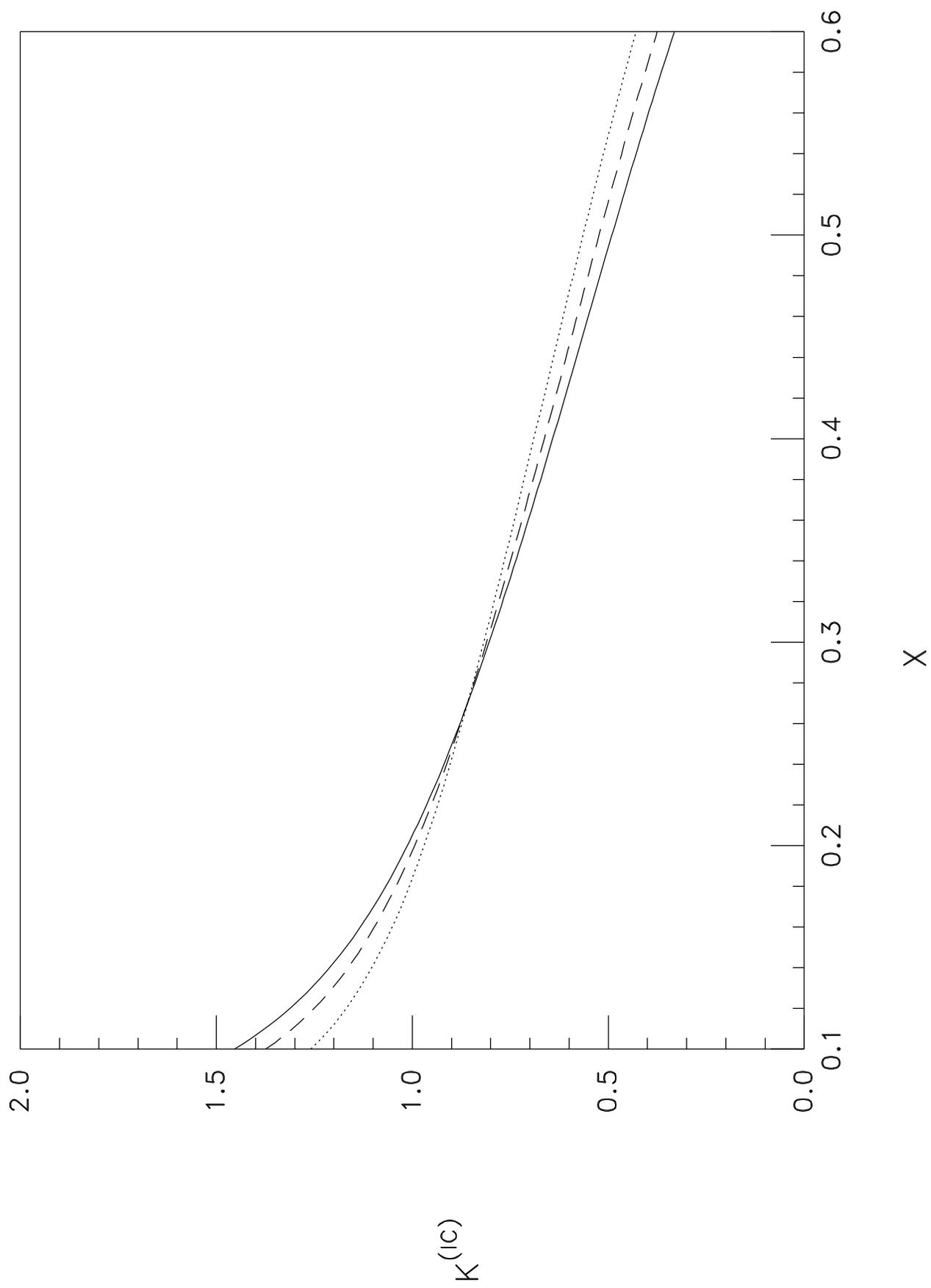
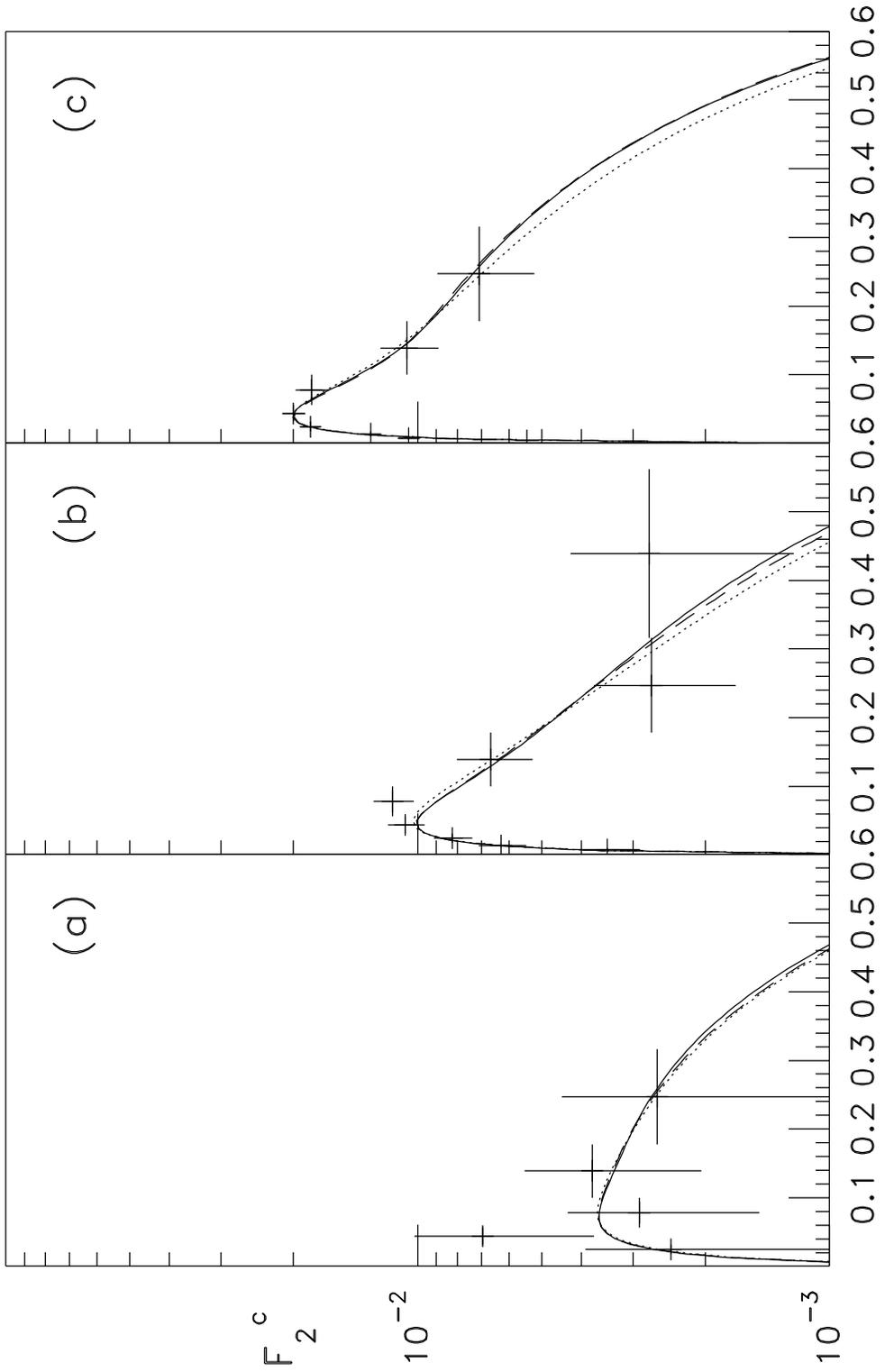
\end{document}